\documentstyle[aps,epsf,prl,multicol]{revtex}
\begin{document}
%\vspace{0.0cm}
\draft
\title{A semiquantal approach to  finite systems of
interacting particles}
\author{F.Borgonovi$^{[a,b,c]}$, G.Celardo$^{[c]}$,
F.M.Izrailev$^{[d]}$ and  G.Casati$^{[e]}$ }
\address{
$^{[a]}$Dipartimento di Matematica e Fisica, Universit\`a Cattolica,
via Trieste 17, 25121 Brescia, Italy  \\
$^{[b]}$Istituto Nazionale di Fisica della Materia, Unit\`a di Brescia,
 Italy\\
$^{[c]}$Istituto Nazionale di Fisica Nucleare, Sezione di Pavia,
via Bassi 6, 27100 Pavia, Italy\\ $^{[d]}$Instituto de F\'isica,
Universidad Aut\'onoma de Puebla, Apdo. Postal J-48, Puebla 72570,
M\'exico\\ $^{[e]}$ International Centre for the Study of Dynamical systems and Istituto Nazionale di Fisica della Materia and INFN, 22100
Como, Italy\\ }
\date{\today}
\maketitle
\begin{abstract}
A novel approach is suggested for the statistical description of quantum
systems of interacting particles. The key point of
this approach is that a typical eigenstate in the
energy representation (shape of eigenstates, SE) has a well
defined classical analog which can be easily obtained from the
classical equations of motion. Therefore, the occupation numbers
for single-particle states can be represented as a convolution of
the classical SE with the quantum occupation number operator for
non-interacting particles. The latter takes into account the
wavefunctions symmetry  and depends on the unperturbed energy spectrum
only. As a result, the distribution of occupation numbers $n_s$
can be numerically found for a very large number of interacting
particles. Using the model of interacting spins we demonstrate
that this approach gives a correct description of $n_s$ even in a
deep quantum region with few single-particle orbitals.
\end{abstract}
\pacs{PACS numbers: 05.45.-a}
\begin{multicols}{2}

In many physical systems such as complex atoms, heavy nuclei and
interacting spins, highly excited eigenstates in the unpertured
many-particles basis can be treated as a chaotic superposition of
a very large number of components, for a relatively strong
interaction among particles (see, e.g.
\cite{FGGK94,zele,spins}). This fact has been used in
\cite{FIC96,FI97} in order to develop a statistical
description of closed systems with a finite number of
Fermi-particles. In particular, it was analytically shown that for
a strong enough interaction, a smooth dependence
of occupation numbers on the energy occurs, which is related to
global properties of chaotic eigenstates.

As is known, the direct numerical computation of excited
eigenstates is strongly restricted due to a very large number of
many-particles states, which grows extremely fast with
the  number of particles. However, the mean values of
occupation numbers turn out to depend on the average shape of chaotic
eigenstates in the unperturbed basis, not on exact, specific values of
their components \cite{FI97}. 

In this Letter we  develop a
novel approach to quantum systems with chaotic behaviour in the
classical limit. This approach takes into account both the chaotic properties
of the classical system, and the specific features of the unperturbed
single-particle spectrum. As a result, one can avoid
diagonalization of Hamiltonian matrices of a huge size which may be practically unfeasible.
This kind of approach can be applied to generic Hamiltonian systems with 
two-body interaction of the type: 

\begin{equation}
H=H_0 + V; \,\,\,\,\,\,\,\,H_0 = \sum_{i=1}^N h_0^i;
\,\,\,\,\,\,\,\, V= \sum_{i=1}^N \sum_{j=i+1}^N V_{i,j}.
\label{H}
\end{equation}

Here $H_0$ describes $N$ {\it non-interacting} particles with
$h_0^i$ as single-particle Hamiltonians, and $V$ stands for a
long-range two-body interaction between the particles. In what
follows, we assume that the single-particle spectrum is determined
by a finite number $L$ of single-particles energies $\epsilon_h$,
$h=1,...,L$; however, the approach is valid for more generic
systems with an infinite spectrum.

The unperturbed Hamiltonian $H_0$ determines the many-particle
states $\left| K\right\rangle =a_{s_1}^{\dagger}\,.\,\,.\,\,.\,
a_{s_N}^{\dagger }\left| 0\right\rangle$ (with $a_{s_j}^{\dagger
}$ , $ a_{s_j}$ as creation-annihilation operators), that form the
basis in which the exact eigenstates of $H$ are
represented. As usual, we assume that the basis is ordered
according to increasing  unperturbed energy values
$E_0=\sum_{h=1}^{L} \epsilon _h $.

The distribution of occupation
numbers (DON) for single-particle states is defined by the
relation,
\begin{equation}
n_E (h) = \sum_{K=1}^{M} \langle K | \hat{n}_h | K \rangle
|\psi_K^E   |^2
\label{nqu}
\end{equation}
where $\hat{n}_h= a_h^{\dagger} a_h$ 
is the occupation number operator giving the
occupation numbers $n_h^K=\langle K | \hat{n}_h | K \rangle$ 
equal to $0$ or $1$ for Fermi-particles, and to $0,1,2,...,N$ for
Bose-particles. These numbers $n_h^K$ indicate how many particles
in a many-particle {\it basis} state $|K\rangle$, occupy a
particular single-particle state $|h\rangle$. Correspondingly, the
occupation numbers $n_E (h)$ give the probability that one of $N$
particles in a many-particle {\it exact} state with the {\it
total} energy $E$, occupies a particular single-particle state
$|h\rangle$. The total number $M$ of many-body states equals
$M=L!/(N!(L-N)!)$ for Fermi and $M=(N+L-1)!/(N!(L-1)!)$ for
Bose-particles.

One should note that while in the above expression for the DON,
the eigenfunctions $\psi_K^E$ depend on the total Hamiltonian $H$,
the term $n_h^K$ depends on the unperturbed spectrum only. This
fact is crucial for our semiquantal approach. Due to the chaotic
structure of exact eigenstates, one can make an average of the DON
over a small energy window $\Delta E$ around the fixed value $E$. This
averaging procedure is similar to that used in the conventional
statistical mechanics developed for systems with a finite number of
particles in  contact with a heat bath, or for isolated systems
of an {\it infinite} number of {\it non-interacting} particles.

Expression (\ref{nqu}) for the mean values $n_E (h)$
can be considerably simplified by introducing the so-called {\it shape of eigenfunctions},
SE, (envelope of eigenstates in energy representation). The form
of the SE has been studied in details both in the model with random 
two-body interaction \cite{FI97}, and in dynamical models of
interacting particles \cite{we,BI00}.
The introduction of the average quantity SE  (thus
neglecting  correlations between different components
$\psi_K^E$) represents the key point of our approach.

We assume that the unperturbed many-body energy spectrum has an
intrinsic degeneracy. This situation is typical for spin systems,
and is more complicated in comparison with those studied before
\cite{we}. Below we show how this difficulty can be overcome.
Let us redefine the state $|K\rangle $ by means of a
indexes pair $|j,l\rangle $, where $j$ labels the ``unperturbed
energy'' $E^0_j$ of the many-body state, while $l=1,...,N_j$
labels its degeneracy $N_j$. If there are $N_0$ different
``unperturbed'' energies, one can write, $
\sum_{K=1}^{M}=\sum_{j=1}^{N_0}
\sum_{l=1}^{N_j}$, therefore, one has,
\begin{equation}
n_E (h) = \sum_{j=1}^{N_0} \sum_{l=1}^{N_j}
\langle j,l  | \hat{n}_h | j,l  \rangle
|\psi_{j,l}^E |^2.
\label{nquan}
\end{equation}

According to standard  definition\cite{we}, the SE is given by:
\begin{equation}
W_E ( E^0_j ) =  \sum_{l=1}^{N_j} |\psi^E_{j,l}|^2.
\label{qsf}
\end{equation}
By substituting $|\psi_{j,l}^E|^2 \simeq
\langle |\psi_{j,l}^E|^2 \rangle_l =  W_E(E^0_j)/N_j$
with $<...>_l$ as an average over $l$, we obtain an approximate
expression for the DON in terms of the SE,
\begin{equation}
n_E (h) =  \sum_{j=1}^{N_0} \frac{1}{N_j}
\sum_{l=1}^{N_j}
 \langle j,l  | \hat{n}_h | j,l  \rangle W_E(E^0_j).
\label{crus}
\end{equation}
Needless to say, if an unperturbed spectrum has no degeneracy,
Eq.(\ref{crus}) can be written in a similar way by taking an
average over a small window of energy around $E^0_j$.

As one can see, expression (\ref{crus}) depends on two terms of 
different nature. The first one, $\langle j,l  | \hat{n}_h | j,l
\rangle$, refers to the unperturbed many-particle spectrum and
reflects  the specific properties of a single-particle spectrum, as
well as quantum features related to  Fermi-Dirac or Bose-Einstein
statistics. In contrast, the second term, $W_E(E^0_j)$, refers to
global properties of  eigenstates and describes   {\it
interaction} effects. Therefore, the basic idea of our 
``semiquantal" approach is to substitute the latter  term (SE) by its
classical analog which  can be easily found from classical equations
of motion.

Classical analogs of the SE have been studied in different models,
see, for example, \cite{we,BI00}). In practice, one has to derive
the distribution $W_E (E_0) = P(H_0 = E_0 | H=E )$ for the
probability to find the unperturbed energy $E_0$ for $H_0$, given
the conserved total energy $E$. This can be obtained by
generating many different initial classical configurations on the
energy surface $H=E$ 
or sampling the $H_0(t)$ values generated by one single trajectory
onto the energy surface
and computing the correspondent distribution
of $H_0=E_0$ \cite{CCGI96}. 
The two procedures have been found to give the same results in the chaotic 
region\cite{BI00}.
In order to facilitate the comparison
with the quantum SE, in our numerical simulations the bin size of
$E_0$ equals  the energy distance between neighbour values of
$E^0_j$. In the same way one can define the classical distribution
of occupation numbers, $ n_E (h) = P(h_0^i =h | H=E )$ (see also
\cite{BI00}). For the quantum-classical comparison, the bin size of $h$
is taken to be equal to the spacing between close single-particle
energy levels $\epsilon_h$.

Let us stress that in this semiquantal approach (SA) it is
possible to study specific systems of, let us say, $1000$
interacting particles occupying $10-20$ single-particle levels.
Surely, one expect this approach to be valid for highly
excited chaotic states in a deep semiclassical region. However, by
direct numerical simulations we have found that the SA gives
correct results even for energy values close to the ground
state.

The model that we have studied, consists of $N$ 3-D interacting
spins placed in a magnetic field $B$ directed along the $z$-axis.
In order to have a proper many-body operator one should require a
coupling between {\it all spins} (not only between neighbors). The
Hamiltonian thus reads,
\begin{equation}
H = B \sum_{i=1}^N S_i^z +  \sum_{i=1}^{N-1} \sum_{j= i+1}^{N}
(J_x S_i^x S_j^x + J_y S_i^y S_j^y + J_z S_i^z S_j^z)
\label{h000}
\end{equation}
Using the well known relations $ S_j^{\pm} = S_j^x \pm i S_j^y $,
one can write,
\begin{equation}
\begin{array}{ll}
H =  \sum\limits_{i=1}^N S_i^z [ B + J_z \sum\limits_j S_j^z ] + & \\
 + \frac{1}{4}(J_x-J_y)  \sum\limits_{i=1}^{N-1} \sum\limits_{j=i+1}^{N}
[ S_i^+ S_j^+ + S_i^- S_j^- ]
+ & \\
\frac{1}{4} (J_x+J_y) \sum\limits_{i=1}^{N-1} \sum\limits_{j=i+1}^{N}
[ S_i^+ S_j^- + S_i^- S_j^+ ]
\label{h00}
\end{array}
\end{equation}

In the mean field approximation, 
we put  $ B + J_z \sum\limits_j S_j^z \simeq const = 1$. 
The interaction can be further
simplified by the particular choice $J_x=-J_y=J$. Thus, our
Hamiltonian $H=H_0 + V$ has the following form,
\begin{equation}
H =  \sum_{i=1}^N S_i^z
 + \frac{J}{2} \sum_{i=1}^{N-1} \sum_{j=i+1}^{N}
[ S_i^+ S_j^+ + S_i^- S_j^- ].
\label{h0}
\end{equation}

For simplicity, the $N$ classical
constants of motion $|\vec{S}_i|$ have been set to 1, therefore,
the only free classical parameters are the total conserved energy
$E$ and the interaction $J$.
The classical model  has been studied in \cite{BI00} and it was 
numerically found to be chaotic and exponentially unstable in a 
wide energy range. More precisely,
in order to have strong chaos, one needs the interaction $J$
between particles to be strong enough.  
A convenient choice is to take 
the interaction strength $J =  1/N$. One should stress that this
situation is the most difficult for theoretical studies, see
discussion in \cite{BI00}.

Quantization follows the standard rules, which gives, $ S^2_i 
=\hbar^2 m (m+1) $ and $ S_i^z = \hbar s $ with $ -m
\leq s \leq m $. The action of creation and
annihilation operators is defined by :
$$ S_i^{\pm} |...,s_i,...\rangle
=
\hbar
\sqrt{ m(m+1) - s_i(s_i
\pm 1)} |...,s_i\pm 1,...\rangle $$

where $ | s_1,...,s_N \rangle$ are the non-symmetrized states
(first quantization states). 

There are $L=2m+1$ single-particle energy levels
$\epsilon_h = -\hbar h $ with $h=-m,...,m$. Therefore, the
unperturbed many-particle energy spectrum consists of a number of
degenerate levels with the spacing equal to $\hbar$. Note, that
both the ground state $E_g = -m N \hbar$ and the upper level $E_u
= m N \hbar$  are non-degenerate. The classical limit is  recovered
when spins are allowed to have any possible orientation, that is
$m\to\infty$, and $\hbar \to 0$.

The choice we have done ($J_x=-J_y=J$) allows  to reduce the
dimension of the Hilbert space by, approximately, one half. This
happens because the operator $V$ in (\ref{h0}) connects only those
unperturbed many-body states that are separated by the spacing
$2\hbar$. In what follows, we consider the subset of the many-body
states containing the ground state. From these states we construct
the completely symmetrized states $ |s_1,..,s_N\rangle^S =
|n_{-m},...,n_{m}\rangle $, where the r.h.s. refers to their
second quantization representation. Note, that for the symmetrized
states the distribution of occupation numbers is expected to be
described, for a large number of particles, by the Bose-Einstein statistics.

Hamiltonian (\ref{h0}), in the second quantization,  can be written as :
\begin{equation}
{\cal{H}} = {\cal{H}}_0 + {\cal{V}};\,\,\,\,\,\,\, {\cal{H}}_0 =
\sum_{h=-m}^{m} \epsilon_h \hat{n}_h
\label{cal0}
\end{equation}
with $$ {\cal{V}} = \sum_{h=-m}^{m-1} \eta_h
\hat{a}^{\dagger}_{h+1}
\hat{a}^{\dagger}_{h+1} \hat{a}_{h} \hat{a}_{h} +
\sum_{h=-m}^{m-2} \xi_h \hat{a}^{\dagger}_{h+2} \hat{a}_{h}.
$$ Here $\hat{n}_h=\hat{a}^{\dagger}_{h} \hat{a}_{h}$, with
$\hat{a}^{\dagger}_{h}$ and $\hat{a}_{h}$  the
creation-annihilation operators satisfying  the standard
relation $[\hat{a}_{h},\hat{a}^{\dagger}_{k}]= \delta_{hk}$. As
to the coefficients $\eta_h$, $\xi_h$, they have quite
complicated expressions; however, they can be easily computed
numerically.

The procedure we have used in our numerical simulations consists
of the following steps: a) compute the classical values $n_E(h)$, and $W_E(E_0)$
as described above; b) compute the quantum values $n_E(h)$
and $W_E(E_0)$ by  diagonalization of the total
Hamiltonian (\ref{cal0}); c) compute $n_E(h)$ and $W_E(E_0)$ by
using the semiquantal approximation according to expression
(\ref{crus}).

\begin{figure}
\epsfxsize 8cm
\epsfbox{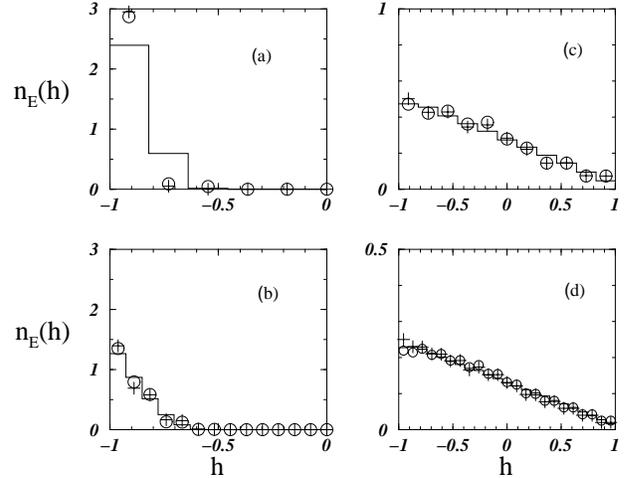}
\narrowtext
\caption{Occupation number distribution obtained in three
different ways. Crosses correspond to the quantum data, open
circles stand for the SA approach, and hystograms refer to the
classical quantities. The data are given for $3$ interacting
particles: (a) $E=-2.75$, $\Delta E = 0.15$, $m=5$, (b)
$E=-2.75$, $\Delta E = 0.15$, $m=13$; (c) $E=-0.91$ $\Delta E
= 0.09$, $m=5$, and (d) $E=-0.91$ $\Delta E = 0.09$, $m=11$.}
\label{n4}
\end{figure}

The results for the DON are summarized in Fig.\ref{n4}. For the convenience of comparison,
symbols refer to the quantum and SA results, while the classical data
are presented as hystograms,.
Note, that each  particle can have  energy within the interval
$[-1,1]$. In order to make the quantum-classical comparison as
close as possible, we took the classical bin size equal to $\hbar$. All
distributions have been normalized in such a way that
$\sum_{h=-m}^{m} n_E (h) = N $.
In cases (a) and (b) we choose the same energy, close to the ground state
and two different $m$ values. In (c) and (d) instead, we choose a higher energy
value. The latter case( (c) and (d)) should describe a 
situation more classical than the former
(higher energy should correspond to higher temperature).
One can see that, while classical and quantum data 
only disagree in the {\it deep} quantum region (a) (energy close
to the ground state and small $m$ ({\it big} $\hbar$)), 
there is a very nice correspondence between quantum and SA data 
in all cases (including region (a)).  
This nice correspondence  is far from trivial
since  SA does not take into account 
quantum correlations
inside exact eigenstates, together with possible
correlations between the two terms, $\langle j,l  |
\hat{n}_h | j,l  \rangle$ and $ |\psi_{j,l}^E |^2$ in
expression (\ref{nquan}), both expected to be strong in the deep 
quantum region.

It is very interesting to explore the occurence of the
Bose-Einstein (BE) distribution in our model. A similar problem has
been studied in detail for the model of two-body random
interaction \cite{FI97} where the conditions for the appearence of
Fermi-Dirac distribution have been found for few interacting
Fermi-particles.

By assuming, a priori, the validity of the BE-distribution
$n_E^{BE} (i) = [e^{\beta (\epsilon_i-\mu)}-1]^{-1}$ in our closed
system, one can find the ``temperature" $1/\beta$ and the
``chemical potential" $\mu$ via  the standard relations,
\begin{equation}
\begin{array}{ll}
\sum\limits_{i=-m}^{m} n_E^{BE} (i) = N;\,\,\,\,\,\,\,
\sum\limits_{i=-m}^{m} n_E^{BE} (i) \epsilon_i = E'.
\label{sys}
\end{array}
\end{equation}
Here $E'$ is the numerically computed energy obtained from the single-particle
quantum distribution (see details in \cite{BGIC98}), and $N$ is the
number of particles. Notice that, due to interaction, $E' \ne E$.

\begin{figure}
\epsfxsize 8cm
\epsfbox{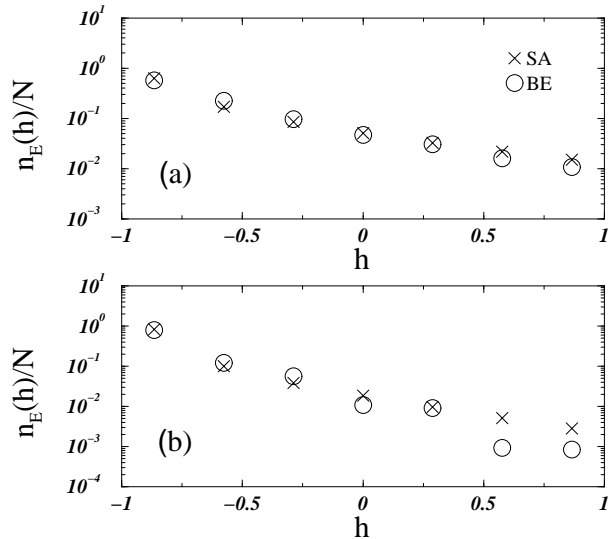}
\narrowtext
\caption{Distribution of occupation numbers for $N=10$ particles.
Open circles indicates the BE-distribution while crosses stand for
the SA data: (a) $E=-8$, $\Delta E = 0.2$, $m=3$, (b)
$E=-6.5$, $\Delta E = 0.2$, $m=13$.}
\label{n10}
\end{figure}

The comparison between BE and SA distributions 
are shown in Fig.2. As one
can see, even for relatively small $N=10$, the distribution
$n_E(h)$ can be closely approximated by  BE-distribution.
This  confirms the
expectation that a strong enough  interaction between
particles can play the role of an internal heat bath \cite{FI97}.
Therefore, the standard quantum 
distributions can be used, with a corresponding renormalization of
the energy $E'$ (see details in \cite{FI97}).
Thus in our model there is no need to increase further the number of particles,
since even for $N=10$ a remarkable agreement with BE distribution has been found.

We would like to stress that an exact quantum treatment of the last 
example
calls for a diagonalization of a huge matrix of size
$8008 \times 8008$, while,  with the semiquantal approach,  all
computations required few minutes on a standard portable PC.

In conclusion, we suggest an effective semiquantal approach to
closed systems of interacting particles, based on the chaotic
structure of eigenstates. In this approach, the computation of the
distribution of occupation numbers can be easily performed by
making use of the classical analog of the shape of eigenstates in
the unperturbed many-particle basis. We demonstrate the
effectiveness of this approach using the model of  3-D
spins with anisotropic Ising interaction. The data show that
semiquantal computations give results which are very close to the
exact ones.

\section{Acknowledgments}

FMI acknowledges the support by CONACyT (Mexico) Grant No.
34668-E.

\end{multicols}
\end{document}